# Great Restraining Wall in Multidimensional Collective Variable Space

Zhijun Pan[1], Maodong Li[1], Dechin Chen[1], Yi Isaac Yang[1]

[1] Institute of Systems and Physical Biology, Shenzhen Bay Laboratory, Shenzhen 518132, China

## Abstract

Enhanced sampling methods are pivotal for exploring rare events in molecular dynamics (MD), yet face challenges in high-dimensional collective variable (CV) spaces where exhaustive sampling becomes computationally prohibitive. While techniques like metadynamics (MetaD) and path-CV enable targeted free energy surface (FES) reconstruction, they often struggle with confinement stability, hyperparameter sensitivity, and geometric flexibility. This work introduces the Great Restraining Wall (GW) method, a robust framework for efficient FES sampling within predefined CV subspaces, addressing these limitations through a novel kernel density estimation (KDE)-derived restraining potential.

GW operates by constructing a bias potential that confines sampling to user-defined regions—ranging from multidimensional masks to 1D pathways—via asymptotically half-harmonic barriers. Unlike MetaD variants requiring iterative bias deposition, GW's potential is derived from a cumulative distribution function, ensuring confinement without manual hyperparameter tuning. GW provides a versatile, stable, and efficient framework for targeted FES sampling, particularly beneficial for complex biomolecular systems with intricate CV landscapes. Its integration with existing enhanced sampling protocols opens avenues for studying ligand binding, conformational transitions, and other rare events with unprecedented precision. Future work will explore GW's extension to adaptive regions and machine learning-guided CV discovery.

## Introduction

Nonlinear interactions, multi-scale, and cooperative behaviors [1, 2] are frequently encountered in fields as varied as molecular biology[3, 4], chemical reaction kinetics[5, 6], and condensed matter physics[7, 8] etc. These complex dynamical systems often exhibit metastable states separated by high free energy barriers, making transitions between stable configurations rare events that occur on timescales far beyond the reach of conventional molecular dynamics (MD) simulations. As a result, traditional MD struggles to satisfy the ergodic hypothesis due to its inherent time-scale limitations, leading to incomplete sampling of conformational space within reasonable cost of computational time. To address this critical bottleneck, enhanced sampling methods[9, 10] have emerged as indispensable tools. By strategically biasing the sampling process through adaptive algorithms, these techniques effectively accelerate rare event sampling, enabling accurate characterization of slow transitions and thermodynamic properties in biologically relevant complex systems[11].

Order parameters or reaction coordinates reflect the important dynamic which is buried by random fast noise such as thermal bath, solvent collision and local vibrational modes. Analogous to projecting system dynamics onto a subspace via tracing out environmental fluctuations[12], dimensionality reduction can be achieved through setting up a serials of low-dimensional variables as collective variables (CVs). CVs are usually geometry functions $\boldsymbol{s}(\boldsymbol{R})$ of the atomic coordinates $\boldsymbol{R}$ of the system such as bonds, angles, dihedrals and root-mean-square deviation (RMSD), which can describe the physical behaviour of slow degrees of freedom (DOFs) [13] to guide the sampling process. Notable CV-based techniques include local elevation[14], adaptive umbrella sampling(AUS)[15, 16], self-healing umbrella sampling (SHUS)[17], Gaussian-mixture umbrella sampling method (GAMUS), metadynamics (MetaD)[18], on-the-fly probability enhanced sampling(OPES)[19] and variationally enhanced sampling (VES)[20, 21]. Another category of approaches including replica-exchange molecular dynamics (REMD)[22], accelerated molecular dynamics(aMD)[23, 24] and integrated tempering sampling (ITS)[25, 26] *etc*. are CV-free. These CV-free approaches achieve universal acceleration across all DOFs in the system without discrimination. CV-based enhanced sampling adaptively accelerates the DOFs associated with the CVs, generally resulting in higher sampling efficiency for specific physical properties—such as chemical reactions[27-30], conformational changes[31, 32], phase transitions[33] or ligand binding processes[34-39]—compared to the CV-free approaches.

However, high-dimensional CVs spaces pose significant scalability challenges for such approaches, as sampling efficiency decays exponentially with increasing CV dimensionality. This exponential barrier becomes particularly problematic when the computational focus is narrowed. Given that exhaustive exploration of the full CV space is impractical and computationally infeasible, to circumvent the curse of dimensionality efficient alchemical FES calculation always rely on approaches with multiple layered restraints by confining sampling to relevant subspaces. The restraint potential of each CV is mostly harmonic to reference value or flat-well in cutoff range switching to half-harmonic walls outside called buckets sampling[40]. B. Roux and C. Chipot have developed the PMF-based restraint[41] with adaptive biasing force (ABF)[42, 43] method via spherical coordinates[44] that assume a rectilinear pathway along one-direction.

Though these methods have been successfully used to compute ligand binding free energy in many cases, whereas the bound and unbound states of the ligand are well characterized, the unknown intermediate **path** connecting them remains challenges their broader applicability. To obtain a minimum (free) energy pathway connecting two metastable states, one-dimensional chain-of-states(CoS)[45] methods such as the replica path method (RPath)[46], nudged elastic band (NEB)[47], reaction path with holonomic constraints (RPCons)[48] and constant advance replicas (CAR)[49, 50] method represent the transition path with a series of discrete minimized or equilibrium conformations. Enhanced sampling of one dimensional CoS CV can also be performed using Boxed MD(BXD)[51, 52] as CAR-BXD[50].

Path-CV[53, 54] is another reparametrize[55], self-evolution two-dimensional $(s, z)$ approach similar to chain-of-states methods, however require auxiliary restraining potentials on the orthogonal $z$-axis to restrain sampling within one dimension path $s$. Some other targeted approaches such as funnel-MetaD[56] as a combination of geometric restraints with the metadynamics technique offers a good balance between accuracy and efficiency and is particularly well suited for the study of rigid ligands lying at the surface of a protein. However, for flexible or buried ligands which is nontrivial to select an appropriate set of CVs to describe the relative movement of the ligand with respect to the protein, funnel-MetaD face convergence issues. Therefore, "Sinking" metadynamics(SinkMeta) [57] have been propose by our team and applied to broad range of ligand-binding FES and drug-protein dissociation dynamics [58]. The center of mass (COM) of ligand was restrained to arbitrary defined region with incomplete knowledge of pathway, then success escaping trajectories are sampled even for deeply buried case in the protein pocket.

However, two critical interventions are necessary for SinkMeta: carefully selecting the initial configuration within the target region and precisely adjusting an additional parameter to suppress the maximum value of the biasing potential, thereby preventing escape from the region. This article introduces the Great restraining Wall (GW) method, a SinkMeta-inspired framework for targeted free energy surface (FES) sampling inside predefined regions of CV space, with emphasis on its wall-like behavior outside the region. The GW method's core innovation lies in its kernel density estimation (KDE)-derived cumulative distribution function, which generates universal, MetaD-independent restraining potentials. Unlike SinkMeta—which relies on localized sampling—GW enforces confinement through asymptotically half-harmonic barriers. These barriers inherently restrict CV exploration to predefined zones without hyperparameter tuning, ensuring robust stability and broader applicability across diverse sampling geometries. We demonstrate GW's versatility through multi-dimensional case studies and combination with other enhanced sampling method, reconstructing specific target regions of FES for geometrically diverse CV subspaces with high precision and shorten MD time.

## Methodology

Metadynamics (MetaD) is a powerful computational method designed to enhance the sampling of rare events in molecular simulations by explicitly biasing pre-selected CVs, which serve as low-dimensional representations of complex system states. The technique reconstructs the FES of the system by iteratively adding a history-dependent bias potential $V(\mathbf{s};t)$ in the form of Gaussian functions deposited at regular intervals. These Gaussians "fill" free energy wells, discouraging the system from revisiting already sampled CV values and thus accelerating exploration of the FES in CV space. A recent evolution of MetaD named on-the-fly probability enhanced sampling explore ($OPES_e$)[59] which aim at making the system sample a broadened target probability distribution $p^{tg}(\mathbf{s})$ such as the well-tempered distribution. To achieve this, Gaussian kernels are used to reconstruct $p^{tg}(\mathbf{s})$, which in turn determines the bias potential $V(\mathbf{s};t)$ through a recursive strategy.

The bias potential V($\mathbf{s}$;t) in MetaD/OPES$_e$ is constructed from a sum of Gaussians kernel function[60] $G(\mathbf{s(t)},\mathbf{s}(t'))=\exp\left[-\frac{1}{2}\left\|\frac{\mathbf{s(t)}-\mathbf{s(t')}}{\boldsymbol{\sigma}}\right\|^2\right]$ centered at previously visited CV configurations $\mathbf{s}(t')$ $\{s^{(1)}(\boldsymbol{R}), s^{(2)}(\boldsymbol{R}), \ldots, s^{(D)}(\boldsymbol{R})\}$, with weight $w$ and width $\boldsymbol{\sigma}$. Only a few numbers of Gaussian kernels at the grids $\{\boldsymbol{s}_i\}$, $i$={1,2⋯N}∈ℕ are need to covers the configuration region $\mathbf{\Omega(R)}$ where the system is interested.

Convolutional Metadynamics (ConvMeta)[57] use a novel gridded convolutional approach to estimate adaptive bias potential V($\mathbf{s}$;t):

$$V(\mathbf{s};t) = \sum_i^N k_i(t)\Delta S_i e^{-\frac{1}{2}\left\|\frac{s-s_i}{\sigma'}\right\|^2} \quad (1)$$

Here, $k_i(t) = \sum_t \frac{\omega(t)}{C(t)} e^{-\frac{1}{2}\left\|\frac{s_i - s'(t)}{\sigma'}\right\|^2}$ is the $i$-th time dependent gaussian-form weight parameter, the $i$-th time independent gaussian-form basis functions $g_i(\boldsymbol{s}) = \exp\left[-\frac{1}{2}\left\|\frac{s-s_i}{\sigma'}\right\|^2\right]$. With standard deviation $\boldsymbol{\sigma}' = \boldsymbol{\sigma}/\sqrt{2}$ on the CV-grids $\{\boldsymbol{s}_i\}$ to fit the Gaussian function $G_i(\boldsymbol{s}(\boldsymbol{R}); t)$ with standard deviation $\boldsymbol{\sigma}$.

The $\mathbf{s}_i$ is the discretization grid coordinate, $\mathbf{s}'(t)$ is the CV value as Gaussian hill center at time t to be added to the total bias potential V($\mathbf{s}$;t). The D-dimensional grid bin-width $\Delta \mathbf{S}_i = \prod_d^D \Delta s_i^{(d)}$ is the product of each dimension of the $i$ -th gird spacing $\Delta \mathbf{s}_i = \left\{\Delta s_i^{(1)}, \Delta s_i^{(2)}, \ldots, \Delta s_i^{(D)}\right\}$. Here we use a $D$-dimensional vector as the standard deviation $\boldsymbol{\sigma} = \{\sigma^{(1)}, \sigma^{(2)}, \ldots, \sigma^{(D)}\}$because only the diagonal matrix is used as the covariance matrix in most cases.

Metadynamics always drives the system away from its current CV position due to the positive repulsive bias potential V($\mathbf{s}$;t) . However, the bias potential perpetually accumulates along the CV space without a natural termination condition—unless striking to a boundary 'wall'. SinkMeta simply subtract the weight parameter $k_i(t)$ with a time dependent but CV space universal shift factor to result in a "negative" bias potential $V_{sink}(\mathbf{s};t)$:

$$V_{sink}(\mathbf{s};t) = \sum_i^N (k_i(t) - v_{shift}(t))\Delta S_i e^{-\frac{1}{2}\left\|\frac{s-s_i}{\sigma'}\right\|^2} \quad (2)$$

Comparing Eq. (2) to Eq. (1), the additional bias potential $V_{shift}(\mathbf{s};t)$ corresponding to SinkMeta is proportion to cumulative function $\Phi(\mathbf{s})$:

$$V_{shift}(\mathbf{s};t) \equiv V_{sink}(\mathbf{s};t) - V(\mathbf{s};t) = -v_{shift}(t)\Phi(\mathbf{s}) \quad (3)$$

$$\Phi(\mathbf{s}) = \int_{\Omega} d\xi e^{-\frac{1}{2}\left\|\frac{\xi-\mathbf{s}}{\boldsymbol{\sigma}'}\right\|^2} \approx \sum_i^N \Delta \mathbf{S}_i e^{-\frac{1}{2}\left\|\frac{\mathbf{s}_i-\mathbf{s}}{\boldsymbol{\sigma}'}\right\|^2} \quad (4)$$

Here, the shift factor $v_{shift}(t) = \frac{V_{max}(t)+E_{depth}}{C'}$, where the normalization constant $C' = \Phi(\mathbf{s}(V_{max})) \approx \int_S d\xi e^{-\frac{1}{2}\left\|\frac{\xi-\mathbf{s}}{\boldsymbol{\sigma}'}\right\|^2}$ to make sure the maximum of sinking bias potential is $-E_{depth}$. $V_{max}(t)=\max\{V(\mathbf{s};t)\}$ is the maximum of the origin bias potential before applying the sinking shift.

SinkMeta innovates in trapping systems in arbitrary-shaped target regions $\mathbf{\Omega}(\mathbf{R})$ by slowly increase restraint potentials $V_{shift}(\mathbf{s};t)$ via the multiplying of global scaling factor $v_{shift}(t)$ and cumulative function $\Phi(\mathbf{s})$ at the margin region, see Eq. (3)(4). However, the "stair" disappears at distal region due to the second properties ii of $\Phi(\mathbf{s})$(see figure 1a,1b and Appendix A), there is no way to come back. Therefore, the initial configuration region cannot outside the $\mathbf{\Omega}(\mathbf{R})$ region. What's more, SinkMeta need to adjust the extra parameter $E_{depth}$ carefully to avoid escaping from region $\mathbf{\Omega}(\mathbf{R})$.

To achieve precise control over free-energy convergence while maintaining reversibility between metastable states, the restraint potential should satisfy two demands: (1) dynamically confine systems to target regions , and (2) allow smooth re-entry from distal CV space. In SinkMeta, the first demand was partially addressed via shift potentials $V_{shift}(\mathbf{s};t)$ while an extra parameter $E_{depth}$ have to be tuned. The second demand require a growing wall that steeper with distant from target region instead of the flatten plateau in $V_{shift}(\mathbf{s};t)$ of SinkMeta. To overcome these two drawbacks of SinkMeta, Great restraining Wall (GW) method is proposed by scaling logarithm of cumulative function ln($\Phi(\mathbf{s})$) instead:

$$V_{GW}(\mathbf{s}) = -K\ln(\Phi(\mathbf{s})) + K\ln Z_p \quad (5)$$

The second term in Eq. (5) is a constant potential shift of the restraining wall potential within region $\mathbf{\Omega}$ that make it close to zero: $V_{GW}(\mathbf{s}) \approx 0$, $\mathbf{s} \in \mathbf{\Omega}$. The definition of the normalization factor $Z_p$ is in Appendix A, Eq. (A3).

Two simple prerequisite conditions are required:

1. For each dimension d={1,2⋯D}∈ℕ, full width at half maximum (FWHM) of $\Delta s_i^{(d)} < \sigma^{(d)}$ ensures sufficient Gaussian kernel overlap for a smooth sum, see S-II section in Supporting Information for detail.

2. The equality of grid distance $\Delta s_i^{(d)}$ for each $i$={1,2⋯N}∈ℕ.

Once the two condition are satisfied, the logarithm of cumulative function $\ln(\Phi(\mathbf{s}))$ in GW method will creates two distinct regimes: a plateau in the target region, steep boundary wall as CV $s$ approaches the margin, which yielding infinity as $\Phi(\mathbf{s})\rightarrow 0, \ln(\Phi(\mathbf{s}))\rightarrow -\infty$ . For the case of one-dimensional CV space, restrained region defined as $[s_{min}, s_{\max}]$, GW potential can express as logarithm of the error function difference:

$$V_{GW}(s)=-K\ln\left[\mathrm{erf}\left(\frac{s-s_{min}}{\sigma\sqrt{2}}\right)-\mathrm{erf}\left(\frac{s-s_{max}}{\sigma\sqrt{2}}\right)\right]+K\ln 2 \qquad (6)$$

$$\mathbf{F}_{GW}(s)=\frac{2K}{\sigma\sqrt{2\pi}}\frac{\exp\left[-\frac{1}{2}\left(\frac{s-s_{min}}{\sigma}\right)^2\right]-\exp\left[-\frac{1}{2}\left(\frac{s-s_{max}}{\sigma}\right)^2\right]}{\mathrm{erf}\left(\frac{s-s_{min}}{\sigma\sqrt{2}}\right)-\mathrm{erf}\left(\frac{s-s_{max}}{\sigma\sqrt{2}}\right)} \qquad (7)$$

Here, $\mathbf{F}_{GW}(s)=-\frac{\partial V_{GW}(s)}{\partial s}$ is the force of GW.

The GW asymptotic to half-harmonic wall with spring constant of $\frac{K}{\sigma^2}$ when CV $s$ is far away from region $[s_{min}, s_{max}]$, see Appendix B for details:

$$V_{GW}(s)\approx\begin{cases}\frac{K}{2\sigma^2}(s-s_{min})^2, & s\ll s_{min}\\ \frac{K}{2\sigma^2}(s-s_{max})^2, & s\gg s_{max}\end{cases} \qquad (8)$$

Derivative to s get the force $\mathbf{F}=-\frac{\partial V}{\partial s}$:

$$\mathbf{F}_{GW}(s)\approx\begin{cases}-\frac{K}{\sigma^2}(s-s_{min}), & s\ll s_{min}\\ -\frac{K}{\sigma^2}(s-s_{max}), & s\gg s_{max}\end{cases} \qquad (9)$$

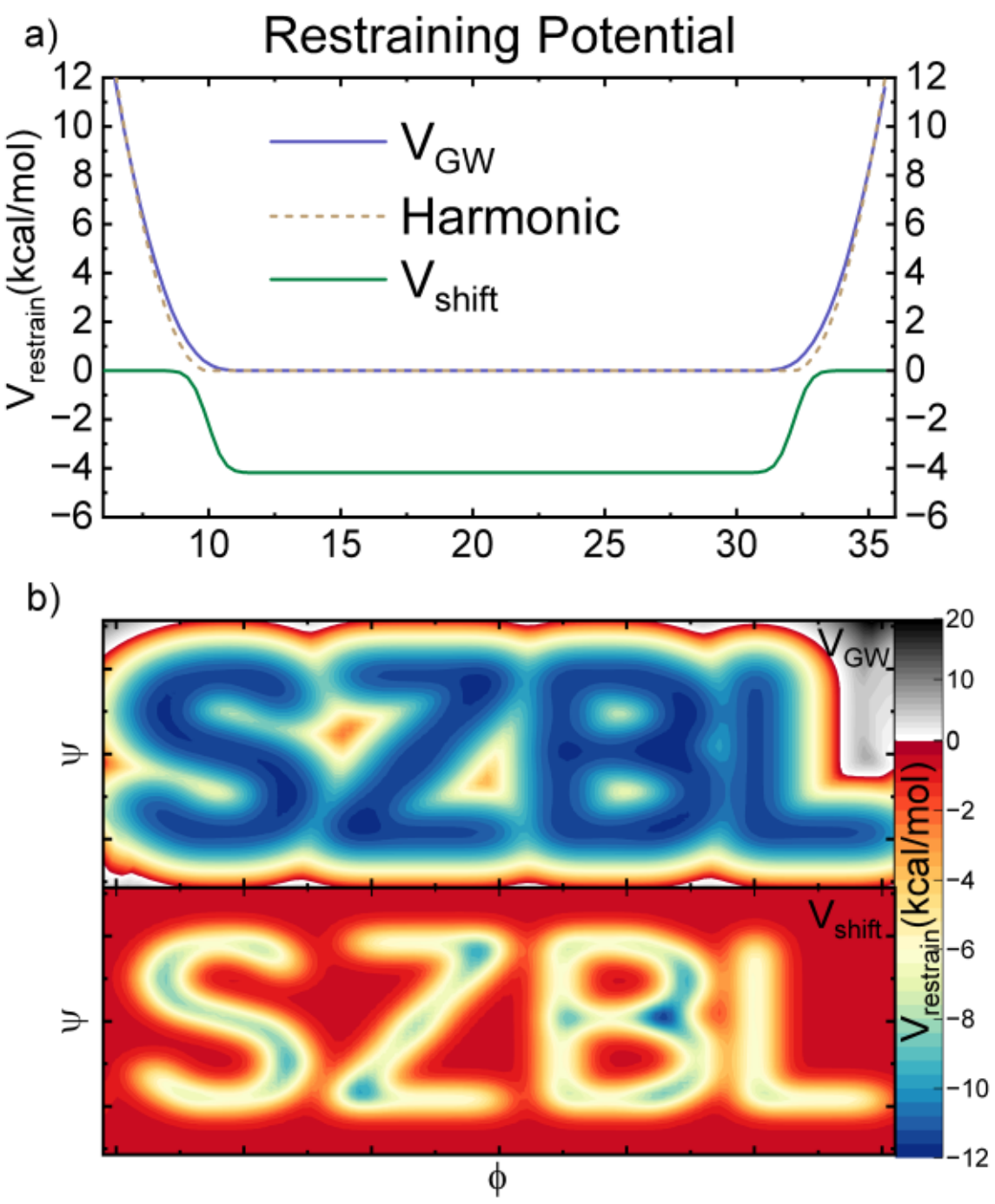

Figure 1. The restraining potential in one- and two-dimensional CV space. a) The 1D-GW biasing potential(blue solid line) vs SinkMeta's shift potential(green solid line), flat-bottom harmonic potential of the same criteria is also plotted in tan dash line as reference. b) 2D-GW biasing potential $V_{GW}$ (upper panel) and shift potential of SinkMeta $V_{shift}$ (lower panel) of the region based on "SZBL" letters in 2D-CV space, the color scales are set to be the same for comparison.

Figure 1 shows some examples of restraint potential in 1D and 2D CV space. For 1D CV space in figure1a), the region of interest is 10-32Å, the Gaussian width is 0.5Å, grid bin width $\Delta s = 0.3$Å; for 2D CV space in figure 1b), the region of "SZBL" letters contains 325 scatter points within a 100X100 grid of CV space (ϕ, ψ). The Gaussian widths are both 0.14 rad, while 2D-grid bin width $\Delta s^{(0)}=\Delta s^{(1)}=0.0628$ rad. More details about the 2D region are demonstrated in the S-I section of Supporting Information.

The novelty of GW method lies in the restraining wall works on versatile configuration region $\mathbf{\Omega}(\mathbf{R})$. For example, it can be clipped into any irregular shape by applying a mask on the original even-space grid. Furthermore, $\mathbf{\Omega}(\mathbf{R})$ can also be reduced to lower dimension subspace: 1D minimum free energy path (MFEP) connecting several metastable states—just like the SinkMeta[57]—merely assume the one-dimensional curve $l$ is parameterized by θ :

$\mathbf{s}(\theta)=\{s^{(1)}(\theta),s^{(2)}(\theta),\ldots,s^{(D)}(\theta)\}$. The line element d**s** is defined via arc length differential:

$$d\mathbf{s}=\left\|\frac{d\mathbf{s}}{d\theta}\right\|d\theta=\sqrt{\sum_{d=1}^{D}\left(\frac{ds^{(d)}(\theta)}{d\theta}\right)^2}\,d\theta \quad (8)$$

Then total line length $L=\int_l d\mathbf{s}$ instead of volume $V$ of the subspace region is divided in the normalization factor calculation:

$$Z_l=\frac{1}{L}\int_l d\mathbf{s}\,\Phi(\mathbf{s}) \quad (9)$$

## Results

Two aqueous solution systems are simulated using CUDASPONGE software: from well-established toy model alanine dipeptide (ALAD) to DNA minor groove binding with a ligand molecules coumarin[36]. In the application of GW, the confinement of region remains robust even when well-tempered bias factor $\gamma\rightarrow\infty$.

### A. ALAD in aqueous solution

We first used a typical alanine dipeptide's water solution as toy system and utilized the two Ramachandran torsion angles $(\phi,\psi)$ as CV space $\mathbf{S}(\mathbf{R})$. This system has four metastable states: C5 (-2.56,2.8), PII (-1.24,2.5), $\alpha_R$ (0.99,0.36) and $\alpha_L$ (0.99,0.36), where C5, PII and $\alpha_R$ are neighboring states in CV space (Figure 2a).

The $\sigma$ of the Gaussian repulsive potential standard deviation and Gaussian kernel in MetaD+GW are all 0.314 rad, and hill height $w=0.4$ kcal/mol. The restrain wall height is also fixed to $K=5.0$kcal/mol. Two kinds of typical regions $\mathbf{\Omega}(\mathbf{R})$ are demonstrated here: 2D irregular area and 1D path as subspace.

1. A 2D mask is applied to clip the full CV space, then we produce a 5ns MetaD+GW with bias factor $\gamma=10$ exploring the masked region. Results are visualized in figure 2:

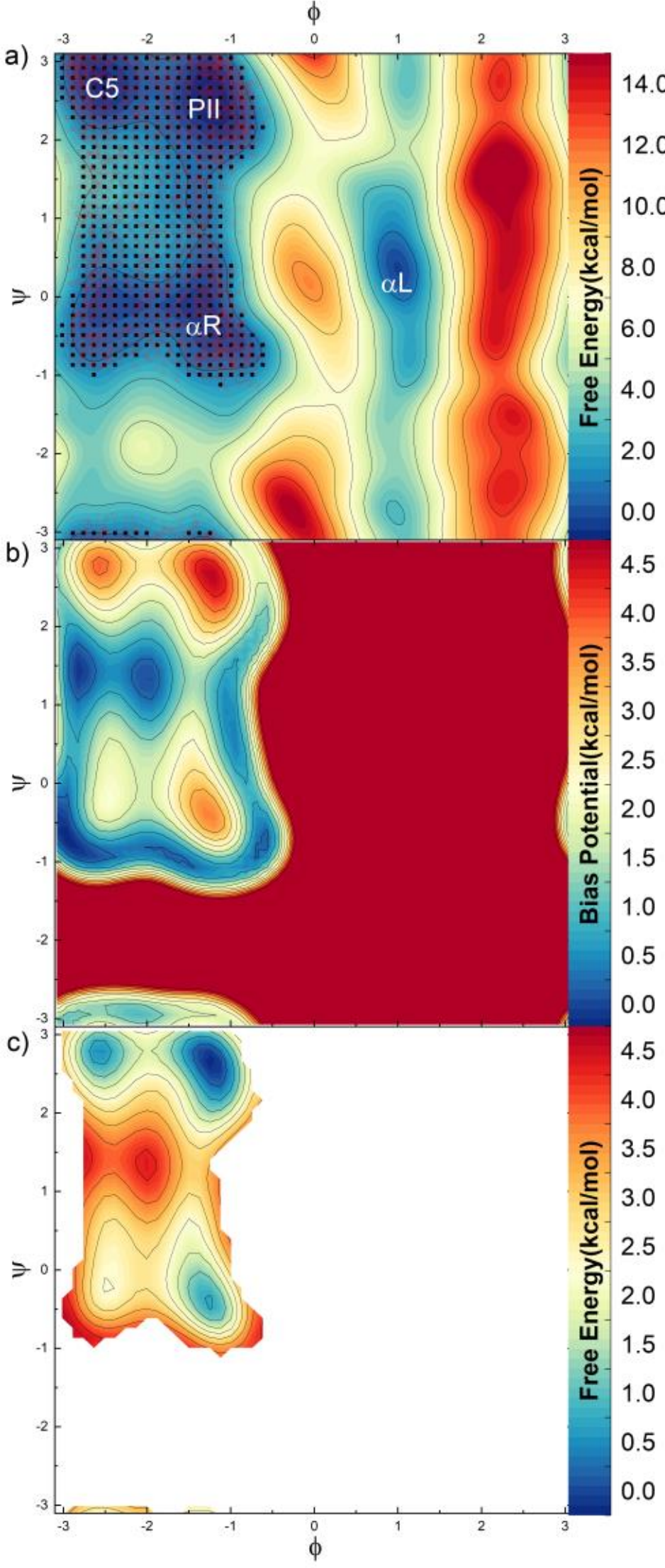

Figure 2. a) FES(kcal/mol) of 2D CV space $(\phi,\psi)$ and the 551-points mask of the local region (black solid square) and the 5ns footprint(red empty circle). b)The bias potential V(**s**) applied to the system in unit of kcal/mol. c) The FES F(**s**) (kcal/mol) calculated only within the masked region.

The figure 2 have shown 2D irregular area sampling within 2D CV space. As illustrated in figure 2a, only 551 Gaussian basis functions are accumulated in areas around C5, PII and $\alpha_R$, see the black square. Figure 2b shows the resulting bias potential as a "great wall" landscape, with steep climbing potential outside the sampling region, see the dense contour lines at the region edge. In fact, the "great wall" potential grows hundred times higher in the distal area where the trajectory(see figure2a's red empty circle) never reaches, thus we choose the color scale to remain linear and consistent to interested region only. Such great wall restrains the sampling to predefined grid mask, thus preventing it from escaping to other areas during the MD simulation. Figure 3c presents the free energy surface calculated using MetaD+GW, demonstrating the successful reconstruction of the local free energy landscape around the C5 PII and $\alpha_R$ states.

2. A closed-loop path was constructed through four dominant local minima C5 (-2.56,2.8), PII (-1.24,2.5), $\alpha_L$ (0.99,0.36) and $\alpha_R$ (0.99,0.36) in the 2D CV space. The path was confined to this 1D elliptical circumference as a subspace. Adopting MetaD+GW with bias factor $\gamma=10$, we produced 1 ns of enhanced sampling, with results shown in figure 3:

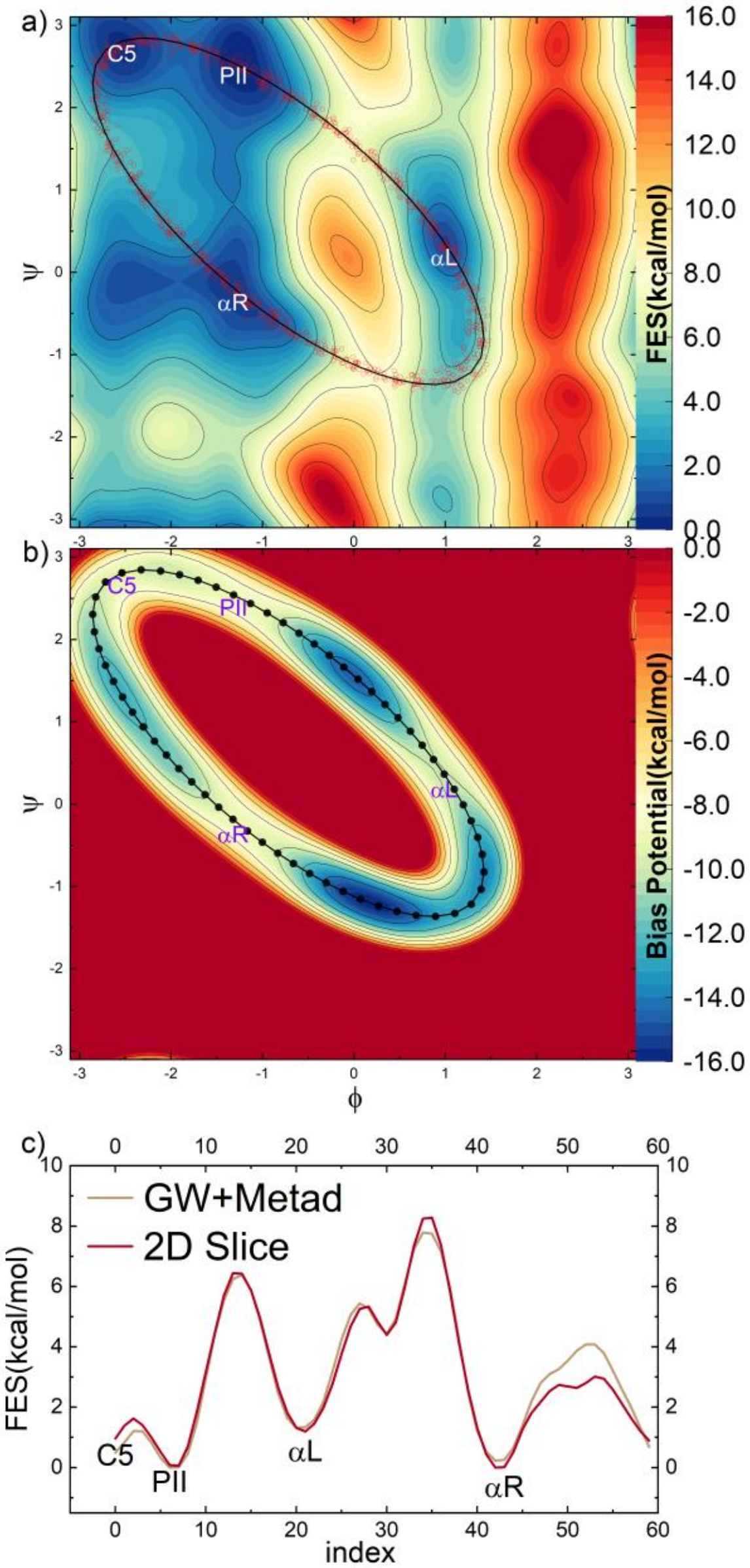

Figure 3. The ellipse path as a close loop crossing 4 dominant local minima: C5, PII, $\alpha_L$ and $\alpha_R$ . a) FES(kcal/mol) of 2D CV space ($\phi$, $\psi$), the 1D ellipse region (black solid line) and 1ns footprint(red empty circle) of MetaD+GW. b)The bias potential V(**s**) applied to the system in unit of kcal/mol and 60 Gaussian kernels' location as black circles with a smooth black line to guide the eye. c) The FES F(**s**) (kcal/mol) calculated only within the ellipse region and the red line is sliced from a) along the ellipse line.

The figure 3 have shown the high efficiency of 1D path sampling in 2D CV space. Only one nanosecond is need to achieve free energy convergence of the interested 4 minima region, while 20ns metadynamics MD is adopted in the reference 2D FES. The accelerate ratio is 20 times.

## B. DNA + coumarin

1D path sampling with MetaD+GW can also be applied to 3D CV space. We demonstrate GW effect in Cartesian space via investigated a DNA-coumarin binding system (DNA duplex sequence d(5'-GCGCATGCTACGCG-3')$_2$), as our SinkMeta method[57] have done before.

We employed MetaD+GW to compute coumarin's relative binding free energies at different sites along the DNA minor groove. The CVs for enhanced sampling was defined as the three-dimensional Cartesian coordinate (x, y, z) of coumarin's center of mass (COM). A serial of 181 Gaussian kernels was placed at $\Delta s = 0.39$Å intervals along the minor groove helix (Figure 4a), centered on the docked coumarin's COM. The MetaD+GW simulations enabled coumarin to diffuse bidirectionally along the DNA's minor groove.

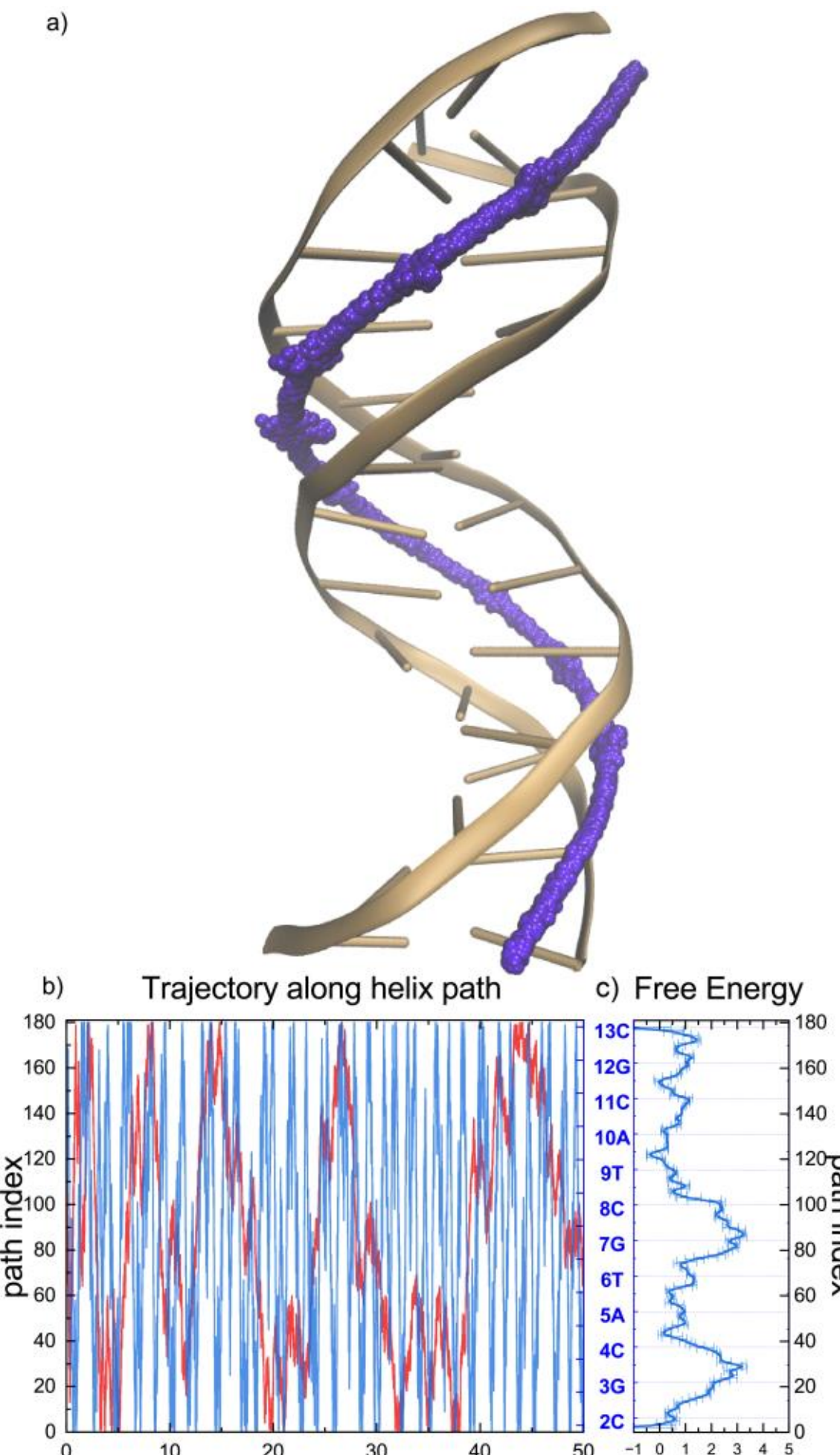

Figure 4: a)50ns MetaD+GW trajectory along minor groove region: Cartesian coordinate trajectory of coumarin's center of mass (purple ball) relative to DNA minor groove. b) MetaD+GW without well-temper (bias factor γ=∞, blue line) compare to bias factor γ=10, SinkMeta in our previous paper[57](red line). c)FES(kcal/mol) calculated along the 181-points path index points.

The minimum relative binding free energy occurs at the site between affinity scores is located between 9T and 10A, predicted by DSDP (see PDB S1) in previous paper, while the next lowest free energy position in Figure 4c is between 4C and 5A corresponding. Both two sites are related to base pair A. This implies that MetaD+GW accurately reveals the relative binding free energy at different DNA minor groove sites. MD simulation with MetaD+GW can also show the molecular details of drug−DNA binding, aiding the investigation of the physiological mechanism of coumarin-DNA interactions.

## C. Apply GW to SITS enhanced sampling

To demonstrate its stability for separate usage, SITS+GW have also been applied to the same system in section A, but using another 2D mask of $\alpha_L$ (0.99,0.36) region as black solid square in figure 5a to clip the full CV space. The temperature ranges from 300-800K with $n_k$=500 as bin for SITS applied on the 22-atoms ACE-ALA-NME dipeptide solute. The iteration phase of have been take out with 10ns. Finally, 10ns production phase with the great restraining wall is run as figure 5b shows the footprint in red empty circles.

The result of SITS+GW shows that GW can be stabilized to the minor metastable state with relatively higher energy. By choosing SITS as another category of enhanced sampling approaches combined the prove the universal of GW can apply. GW focus on achieve higher sampling efficiency while the enhanced sampling methods focus on the acceleration of MD.

The GW potential is everywhere in the full CV space, even far from the interested region. As a result, the green empty circles in figure 5b shows the 1ps relaxation process from initial state $\alpha_R$ to the target $\alpha_L$ region. This demonstrate that necessary to carefully chose the initial configuration within target region in SinkMeta method is overcome by GW method.

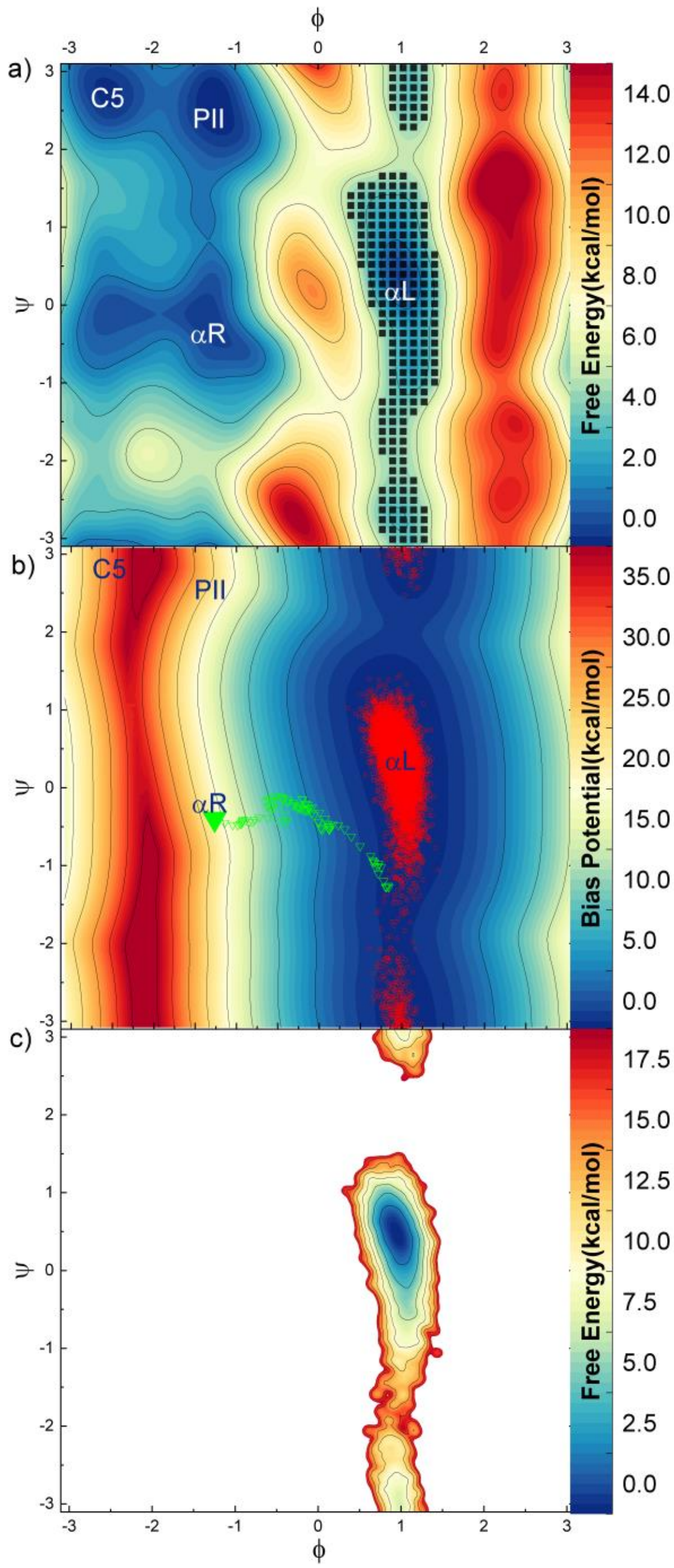

Figure 5. SITS+GW on the 22-atoms ACE-ALA-NME dipeptide solute in water box. a) FES(kcal/mol) of 2D CV space (ϕ, ψ) and 228-points mask of the local region (black solid square) b)The GW potential(kcal/mol) and 1ps relaxation process(green empty triangle) from initial state $\alpha_R$(green solid triangle) to the target $\alpha_L$ region. The total 10ns footprint(red empty circle) is also plot with time interval of 1ps. c) The FES of SITS with a 228-points restraining region generated by reweighting method of PLUMED2[61] using the 10ns footprint(red empty circle) in b).

## Discussion

This approach is a kernel density estimation (KDE)-derived cumulative distribution function extracted from SinkMeta[57], which generalized a universal restraining potential that can be use separately. GW is also more stable than SinkMeta. SinkMeta only restrict to localized sampling: the original potential shift $V_{shift}(\mathbf{s}) \propto \Phi(\mathbf{s})$ proportion to cumulative distribution function damp into simple zero at the far away region, thus an additional hyperparameter $E_{depth}$ is need. Instead, the potential of GW confine CV exploration to user-specified zones via asymptotically half-harmonic barriers outside designated regions, eliminating manual restraint tuning. What's more, since the restraining wall grow to infinity as soon as CV escape from the region edge, the confinement remains robust even for original MetaD(well-tempered bias factor $\gamma \rightarrow \infty$)!

The second case of Result's subsection A demonstrate that GW can samples 1D ellipse path in 2D dihedral angles ($\phi$, $\psi$) as CV space of ALAD in aqueous solution. GW enable a novel continuous sampling strategy which seamless transitions along a defined pathway of CV space. Unlike Path-CV—a two-dimensional approach requiring auxiliary restraining potentials along the orthogonal $z$-axis to constrain sampling to one dimension—GW operates natively in the original CV space.

The two endpoints of path segment in path-CV are not supposed to overlap, which fail to handle periodic pathways such as diagonal line in 2D CV space (ϕ, ψ) (see our SinkMeta paper's figure 4 lower panel). What's more, due to restrain applying to the end along the path is not accurate for FES calculation, the method often need to extended for extra segment. This can be overcome by GW by 1D ellipse case eliminates the demands for unstable restraints on two ends of the path.

By directly exploring CVs along a 1D path, GW achieves efficient free energy landscape estimation with markedly shorter simulation durations compared to multidimensional methods.

DNA is a pharmacological target for many drugs, and its binding to small molecules has multiple modes. Compounds binding to the DNA minor groove has potentialclinical utility against diseases like cancer and parasitic infections[37]. A natural product, coumarin, can bind to the DNA minor groove with various pharmacological properties.

## Conclusion

In this article, the Great Restraining Wall (GW) method ensures a flat energy landscape within the interested region and harmonic confinement outside it is supposed. GW generalizes SinkMeta's concept of enforcing confinement through a robust and versatile intrinsic geometric restraints.

Two system have been applied to demonstrate GW's efficacy. For alanine dipeptide in 2D CV space, GW with a 551-points mask reconstructed the FES and 1D elliptical path connecting metastable states further showcased GW's ability to sample periodic pathways without auxiliary restraints. For a DNA-coumarin system, GW confined sampling to a 3D binding pocket, achieving precise FES estimation despite high bias factors.

GW's restraining potential is MetaD-independent, ensuring stability and eliminating bias decay concerns. What's more, GW supports closed paths directly and inflated geometries (e.g., ligand binding pockets) by adjusting gaussian kernel widths, whereas Path-CV requires orthogonal restraints and fails for periodic paths.

In conclusion, GW provides a versatile, stable, and efficient framework for targeted FES sampling, particularly beneficial for complex biomolecular systems with intricate CV landscapes. Further integration with existing enhanced sampling protocols such as SITS and OPES opens avenues for studying ligand binding and conformational transitions. Future work will explore GW's extension to adaptive regions and machine learning-guided CV discovery.

## Acknowlegments

The authors thank Yi Qin Gao, Xu Han, Yijie Xia, Haohao Fu and Lizhe Zhu for useful discussion. Computational resources were supported by Shenzhen Bay Laboratory supercomputing centre. This work was supported by the National Science and Technology Major Project (No. 2022ZD0115003) and the National Natural Science Foundation of China (22273061 to Y.I.Y).

## Appendix A: Target distribution as additional restraining wall

EDM[62], OPES[19], VES[20] achieve target distribution as soon as these methods reaches convergence when the following additional bias potential is apply to system:

$$V(\mathbf{s})=-F(\mathbf{s})-\frac{1}{\beta}\ln p^{\mathrm{target}}(\mathbf{s}) \qquad \text{(A1)}$$

The second term defines the additional restraining wall proportion to logarithm of target distribution:

$$V_{\mathrm{restrain}}(\mathbf{s})\equiv-\frac{1}{\beta}\ln p^{\mathrm{target}}(\mathbf{s})=-\frac{1}{\beta}\,ln\left[\frac{\Phi(\mathbf{s})}{Z_{\mathrm{p}}}\right]^{\alpha} \qquad \text{(A2)}$$

Properties of $\Phi(\mathbf{s})$:

i. The inner region $\Phi(\mathbf{s})\approx C>0$ is constant.

ii. The distal region $\Phi(\mathbf{s})\approx 0$.

iii. Differentiable: $\frac{\partial\Phi(\mathbf{s})}{\partial\mathbf{s}}=\frac{\mathbf{\Sigma}^{-1}}{\sqrt{2\pi}}\,e^{-\frac{1}{2}\left\|\frac{\xi-\mathbf{s}}{\boldsymbol{\sigma}}\right\|^{2}}$ is Gaussian function, $\mathbf{\Sigma}$ is the covariance matrix of the multivariate Gaussian function, $\mathbf{\Sigma}^{-1}$ is its inverse.

iv. Integral over predefined region is a finite constant:

$$Z_{\mathrm{p}}=\frac{1}{|\mathbf{\Omega}|}\int_{\mathbf{\Omega}}d\mathbf{s}\,\Phi(\mathbf{s}) \qquad \text{(A3)}$$

Here, $Z_{\mathrm{p}}$ is called normalization factor. In the OPES[19] paper Eq. (7), step-dependent normalization factor $Z_n=\frac{1}{|\boldsymbol{\Omega}_n|}\int_{\boldsymbol{\Omega}_n}d\boldsymbol{s}\,\tilde{p}_n(\boldsymbol{s})$. Normalized space was divided by the CV space actually explored at recent step $n$ rather than the full CV space . In our GW case, a fixed volume of *predefined* region $\mathbf{\Omega}$ will divide into the normalization factor $Z_{\mathrm{p}}$ instead.

To construct the corresponding target distribution in the great restraining wall case, start from cumulative function's property then doing some reverse inference. The normalized cumulative function $\Phi(\mathbf{s})/Z_{\mathrm{p}}$ *raised to the power of* $\alpha$ as target distribution will be sampled.

Prof of additional restraining potential of GW:

$$V_{\mathrm{GW}}(\mathbf{s})=-\frac{1}{\beta}\ln\left[\frac{\Phi(\mathbf{s})}{Z_{\mathrm{p}}}\right]^{\alpha}=-\frac{\alpha}{\beta}\ln\left(\frac{\Phi(\mathbf{s})}{Z_{\mathrm{p}}}\right) \qquad \text{(A4)}$$

Now focus on cumulative function,

$$V_{\mathrm{GW}}(\mathbf{s})=-\frac{\alpha}{\beta}\ln\big(\Phi(\mathbf{s})\big)+C \qquad \text{(A5)}$$

Using the relation $\ln(\Phi(\mathbf{s})/Z_{\mathrm{p}})=\ln\Phi(\mathbf{s})-\ln Z_{\mathrm{p}}$ , a constant $C=\frac{\alpha}{\beta}\ln Z_{\mathrm{p}}$ is shifted potential as the result of normalization factor $Z_{\mathrm{p}}$.

Since the derivative of constant is zero, any form of the constant C contributes nothing to restraining force

$\mathbf{F}_{GW}(\mathbf{s})=-\frac{\partial V(s)}{\partial s}$, do not affect the dynamic. However, the C is needed to keep the restraining potential close to zero $V_{GW}(\mathbf{s})\approx 0$ within the region.

Finally, define $K=\frac{\alpha}{\beta}$ as the restraining wall height in unit of energy, Eq. (5) is proved.

For probability-based methods such as OPES[19], the default target distribution is the well-tempered version:

$$p^{target}(\mathbf{s}) \propto [P(\mathbf{s})]^{\frac{1}{\gamma}} \qquad (A6)$$

, where unbiased probability density defined as $P(\mathbf{s})=\langle\delta[\mathbf{s}-\mathbf{s}(\mathbf{R})]\rangle$, bias factor $\gamma>1$ is the well-tempered target ,when in the limit of $\gamma\rightarrow\infty$, $p^{target}(\mathbf{s})$ become a unitary flat target over the full CV space **S**.

Instead of the CV space **S(R)**, the GW focus on the pre-defined region **Ω(R)** only. Within region **Ω(R)** is a flat target distribution while quickly damped to zero outside the region:

$$p^{target}(\mathbf{s}) \propto [\Phi(\mathbf{s})]^{\alpha}$$

Power factor $\alpha>0$ is related to the restraining wall parameter $K=\frac{\alpha}{\beta}$, when in the limit of $\alpha\rightarrow 0$, it becomes a unitary flat target over the full CV space **S**, restraining result trend to the OPES limit of $\gamma\rightarrow\infty$.

## Appendix B: Asymptotic expansion of cumulative function

From Eq. (6), cumulative function is combination of the error function erf(x), the infinity behavior of erf(x) can be described by its asymptotic expansion:

$$\mathrm{erf}(x)=1-\frac{e^{-x^2}}{\sqrt{\pi}x}\cdot\left(1+O(1/x^2)\right),\ x\rightarrow+\infty \qquad (B1)$$

then, at the $s\gg s_{max}$ region Eq. (6) becomes:

$$\begin{aligned} V_{GW}(s)&=-K\ln\left[\mathrm{erf}\left(\frac{s-s_{min}}{\sigma\sqrt{2}}\right)-\mathrm{erf}\left(\frac{s-s_{max}}{\sigma\sqrt{2}}\right)\right]+C \\ &\approx -K\ln\left[\frac{e^{-\left(\frac{s-s_{max}}{\sigma\sqrt{2}}\right)^2}}{\sqrt{\pi}\frac{s-s_{max}}{\sigma\sqrt{2}}}-\frac{e^{-\left(\frac{s-s_{min}}{\sigma\sqrt{2}}\right)^2}}{\sqrt{\pi}\frac{s-s_{min}}{\sigma\sqrt{2}}}\right] \\ &=-K\ln\frac{e^{-\left(\frac{s-s_{max}}{\sigma\sqrt{2}}\right)^2}}{\sqrt{\pi}\frac{s-s_{max}}{\sigma\sqrt{2}}}\left[1-\frac{s-s_{max}}{s-s_{min}}e^{-(2s-s_{max}-s_{min})(s_{max}-s_{min})}\right] \end{aligned}$$

At the $s\gg s_{max}>s_{min}$ condition, the first term in square bracket become dominate, so the latter term is omitted:

$$V_{GW}(s)\approx -K\ln e^{-\left(\frac{s-s_{max}}{\sigma\sqrt{2}}\right)^2}+K\ln\sqrt{\pi}\frac{s-s_{max}}{\sigma\sqrt{2}}$$

At the $s\gg s_{max}$ condition, the first term become dominate, so the latter term is omitted, again.

$$V_{GW}(s)\approx -K\ln e^{-\left(\frac{s-s_{max}}{\sigma\sqrt{2}}\right)^2}=\frac{K}{2\sigma^2}(s-s_{max})^2 \qquad (B2)$$

Vice versa, for $s\ll s_{min}$ limit, the terms relative to $s-s_{max}$ is omitted, finally the asymptotic approximation of the restraining potential at the far away region becomes:

$$V_{GW}(s)\approx\begin{cases}\frac{K}{2\sigma^2}(s-s_{min})^2, & s\ll s_{min}\\ \frac{K}{2\sigma^2}(s-s_{max})^2, & s\gg s_{max}\end{cases} \qquad (B3)$$

Derivative to $s$ get the force $F=-\frac{\partial V}{\partial s}$:

$$F_{GW}(s)\approx\begin{cases}-\frac{K}{\sigma^2}(s-s_{min}), & s\ll s_{min}\\ -\frac{K}{\sigma^2}(s-s_{max}), & s\gg s_{max}\end{cases} \qquad (B4)$$